\documentclass[reprint,prb,letterpaper,amsfonts,amssymb,amsmath,showpacs]{revtex4-1}%
\usepackage{amsfonts}
\usepackage{amsmath}
\usepackage{amssymb}
\usepackage{graphicx}
\usepackage[dvipsnames]{color}
\usepackage{adjustbox}
\usepackage{multirow}

\providecommand\muB{$\mu_\textrm{B}$ }

\begin{document}

\title{Origin of the electron disproportionation in the metallic sodium cobaltates}
\author{Y.\,V.\,Lysogorskiy, S.\,A.\,Krivenko, I.\,R.\,Mukhamedshin, I.\,F.\,Gilmutdinov, O.\,V.\,Nedopekin, and D.\,A.\,Tayurskii}
\email{yura.lysogorskii@gmail.com}
\affiliation{Institute of Physics, Kazan Federal University, Kremlyovskaya St.~16a, 420008 Kazan, Russia}
\date{\today}
	\received[Received ]{date}
	\published[published ]{date}
	
\begin{abstract}

Recently the unusual metallic state with a substantially non-uniform distribution of a charge and mag\-ne\-tic density in CoO$_2$ planes was found experimentally in the Na$_x$CoO$_2$ compound with $x>0.6$.
We have investigated an origin of such electron disproportionation in the lamellar sodium cobaltates by calculating the ion states as a function of a strength of the electron correlations in the $d$(Co)-shells within the GGA+U approximation for the system with a realistic crystal structure.
It was found that the nonuniformity of spin and charge densities are induced by an ordering of the sodium cations and enhanced correlations.
Two important magnetic states of cobalt lattice competing with each other at realistic values of the correlation parameter were found~---~low spin hexagons (LS) and higher spin kagom´e lattice (HS-KSL).
In the heterogeneous me\-tal\-lic  HS-KSL phase magnetic Co ions form a kagom\'e structure. In LS phase the kagom\'e pattern is decomposed into hexagons and Co ions possess minimal values of their spin.
Coexistence of these states could explain the emergence of the disproportionation with the peculiar kagom\'e structure experimentally revealed in previous studies of the cobaltates.
\end{abstract}
	
\maketitle

\section{Introduction}

Mechanisms of correlations between $d$-electrons, hopping in lattices of the transition metal (TM) ions, are crucible for complex many-particle phenomena in the TM-compounds: the insulator-to-metal transition (IMT), high-temperature superconductivity, colossal magneto-resistance, and charge/magnetic ordering.\cite{Imada1998}
Metallic systems with the strong correlations are conventionally obtained from either Mott or charge-transfer insulators, adding a small amount of doped charge carriers in their lattices, as in the case of cuprates and manganites.
But in some cases, the strongly-correlated metallic states arise in doped band insulators, as in ruthenates and iron pnictides/chalcogenides, representing puzzles for the researches.~\cite{georges2013}

A spectacular example of such systems with unconventional correlations is the lamellar cobaltates Na$_{x}$CoO$_2$, composed of CoO$_2$ layers with Na$^+$ ions between them. The layers, containing triangular cobalt lattices, are formed by the edge-sharing CoO$_6$ octahedra slightly compressed along the crystalline $c$-axis.~\cite{wang2003spin,foo2004charge,bernhard2004charge,bayrakci2005magnetic,qian2006quasiparticle,shimojima2006angle}
Na$_1$CoO$_2$ is the band insulator with the spinless Co$^{3+}$ ions.\cite{lang2005evidence}
When the sodium content is decreased in the compound, a metallic state with a rather good electric conductivity is evidenced within the entire range of $x$ except $x=1/2$.

In the range between $x=1/2$ and $x=1$, a simplified intuitive considerations suggest that, when the hole becomes localized at the 3d shells of the cobalt ions, the corresponding $d^5$ magnetic states with $S=1/2$ should appear instead of the spinless $d^6$ state (Co$^{3+}$).
Then in the lattice the proportion $1-x$\,:\,$x$ between the Co$^{3+}$ and Co$^{4+}$ states should exists for any given value of $x$.
However experiments unambiguously indicate that such a picture of the Co$^{3+}$/Co$^{4+}$ segregation is invalid in the cobaltates.
On the contrary, a very peculiar charge/magnetic state called the ``charge disproportionation'' appears.\cite{AlloulEPL2008,mukhamedshin2015order}
In particular, in the doped CoO$_2$ layers a small amount ($\leq25\%$) of Co1 ions remains in the localized non-magnetic state Co$^{3+}$, whereas the holes are delocalized over the remaining part of the cobalt lattice, the Co2 sites.
In the range $0.65<x\lesssim0.80$ at least four stable phases $x=2/3, 0.71, 0.72, 0.77$ exist in the system, which are distinguished by the sodium atom arrangements.\cite{AlloulEPL2008}
Experiments exhibit a well defined correlation between the disproportionation of the $d$-electrons and the ordering of the ions in the sodium layers.\cite{mukhamedshin2015order}
Both phenomena emerge concurrently, disappearing at temperatures higher than $T^{\ast}\approx400~\text{K}$.
The magnetic susceptibility of the system with the disproportionation is temperature dependent, resembling the Curie-Weiss law relevant for localized magnetic moments.
Above 100\,K the dependencies $\chi(T)$ look very much alike for all phases, whereas at smaller temperatures they are different.\cite{AlloulEPL2008}
Namely, upon decreasing $T$ down to $50~\text{mK}$, the phases with $x<0.75$ remain paramagnetic with inplane/interplane ferromagnetic (FM)/antiferromagnetic (AFM) correlations between $d$-moments, while the $x=0.77$ phase exhibits a transition to the A-AFM order at $T_N=22~\text{K}$.

\begin{figure*}[t!]
	\centering
	\includegraphics[width=0.95\linewidth]{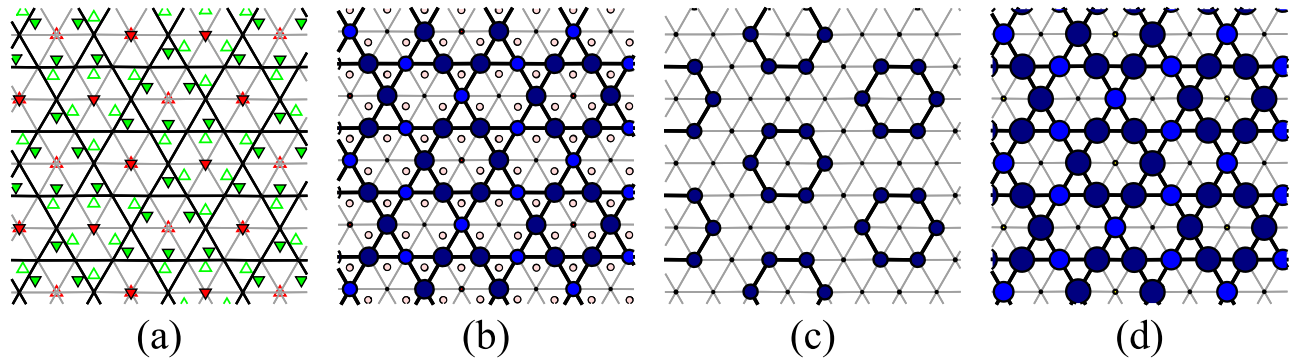}\\
	\caption {\label{fig:Kagome} (Color online) Structure of the lamellar cobaltate Na$_{2/3}$CoO$_2$. The crystal axis $c$ is directed orthogonally to the figure. The solid lines connect the nearest sites of the triangular Co sublattice in the CoO$_2$ layers. Co2 sites form kagom\'e sublattice are highlighted by the thick solid lines. (a) Ordering of the Na$^+$ ions above (filled symbols) and below (open symbols) the CoO$_2$ layer. Na1 (Na2) atoms are presented by red (green) triangles.\footnote{In the sodium layers the Na2 ions form the triangular clusters, containing six adjacent atoms, where the apex positions are conventionally denoted as Na2a sites, and the remaining Na2 positions are denoted as Na2b sites.} (b) Experimentally established disproportionated state of the $d$-electrons.~\cite{AlloulEPL2009,Platova2009} The magnetic Co2a and Co2b ions are given by the light-blue and dark-blue circles, respectively. Pink circles represent the oxygen atoms in the layer just above the layer. (c) and (d) \emph{Ab-initio}  calculated electronic states with $U=5~\text{eV}$: (c) hexagonal LS phase, (d) HS-KSL phase. Circle size is proportional to the magnetic moment at cobalt ion.
}
\end{figure*}

The $x=2/3$ crystal phase was investigated in greater detail than the other ones.
Its three dimensional structure has been reliably established employing both the NMR/NQR and x-ray diffraction techniques.\cite{AlloulEPL2009,Platova2009}
An elementary cell of this phase comprises $264$ atoms.
In particular, the cell has six sodium layers, alternating with six CoO$_2$ layers.
In the sodium planes the ions occupy the two positions: Na1 sites located just above and below the Co1 ions, and Na2 sites being more distant from the nearest Co2 ions.
A sodium layer pattern contains the lone Na1 ions interlaced with six clustered Na2 ions forming triangles, see Fig.~\ref{fig:Kagome}(a).
In the cobalt plane the holes are pushed out from the Co1a and Co1b sites; this makes these ions non-magnetic.
These itinerant holes are delocalized over the Co2 sites making them magnetic. Co2 sites bearing magnetic moments form a regular kagom\'e sublattice (KSL), see Fig.~\ref{fig:Kagome}(b).\cite{AlloulEPL2009}
To be more precise, the KSL contains the less magnetic and more magnetic  Co2a and Co2b ions, respectively.
The $^{59}$Co NMR measurements detected a strongly pronounced in-plane anisotropy of the $d$(Co2)-shells of the Co2 sites.\cite{Platova2009}
This implies that the KSL is formed concurrently by a redistribution of the holes from the axially-symmetric $a_{1g}$ orbitals to other $d$-states of the Co2 sites.\cite{AlloulEPL2009,Platova2009}
The origin of such intricate state of the CoO$_2$ layers remains unclear.

In previous studies of the cobaltates, several theoretical approaches to the problem of their unusual correlated state were developed.
As a first approximation, the interaction of the Co ions with the ligands splits their $3d$ states into the lower $t_{2g}$-triplets separated from the upper $e_{g}$-doublets by the large gap with the energy $\Delta\approx2~\text{eV}$.\cite{Singh2000}
The ``minimal'' approach treats the ground-states configurations of the ions, $t^6_{2g}$ with $S=0$ and $t^5_{2g}$ with $S=1/2$, as being the low spin (LS) states of the sites Co$^{3+}$ and holes Co$^{4+}$, respectively.
In this case, the relevant interaction between the itinerant holes comes from their mutual repulsion within the ion $t_{2g}$ shells with the corresponding Hubbard energy $U$.\cite{
	MaekawaKoshibae2003,baskaran2005ice,
	Zhang2004,Zhou2005,Kotliar2007,Korshunov2007,Piefke2010,Peil2011,Boehnke2012,
	Lee2004,Li2005,Indergand2005,Lee2005,Lee2006,
	lysogorskiy2012ab,Wilhelm2015}
For the temperature greater than a few hundred Kelvins, when the sodium layers are disordered, the correlation effects in the cobaltates are well described by the Hubbard model for the holes in the triangular $d$-lattice.\cite{Piefke2010,Boehnke2012,Wilhelm2015}
At the very high temperature $T\approx 500~\text{K}$ the model reveals some tendency towards spatially non-uniform state of the holes at $x\approx 0.7$, with a weak redistribution of their density over the sites possessing a symmetry of the KSL.\cite{Boehnke2012}
Unfortunately the theory of Ref.~\onlinecite{Boehnke2012}
does not take into account the patterning of Na$^+$ cations, which arise concurrently with the charge redistribution in the Co lattice upon decreasing $T$.\cite{AlloulEPL2009,Platova2009}
The non-uniform Coulomb potential introduced by the sodium layers could substantially affect the hole states in the cobaltates.\cite{roger2007patterning,Kotliar2007,Peil2011}
The disproportionation \emph{per se}, when the holes are almost completely pushed out from the Co1 sites, was not achieved in this model.
Besides, the employed $a_{1g}$ states of the holes\cite{Boehnke2012} do not produce the anisotropy of the $d$-shells, which was observed experimentally.
Again, the Hubbard model does not explain why the KSL disproportionation, inherent to the sodium cobaltates, is not observed in other systems with the triangular $d$-lattices, e.g., the metallic lithium cobaltates Li$_x$CoO$_2$.

The approach used in Refs.~\onlinecite{Khaliullin2005,daghofer2006magnetic,chaloupka2007spin,Chaloupka2008,khaliullin2008origin} suggests that the hole could hop from its site to the nearest $d$-sites, when the counter-directional electron hops back to the $e_g$ orbital.
At the original site this results in the excitation of the Co$^{3+}$ ion to its higher spin (HS) state $t_{2g}^5e_g$ with $S=1$.\footnote{
The $S=1$ and $S=2$ states of the Co$^{3+}$ ion are conventionally called as the intermediate and high spin states, respectively.
In the present study the non-LS states of the $d$-ions are called as the higher-spin states.
However, the $S=2$ states of the Co$^{3+}$ ions do not participate in the hopping of the holes in the idealized CoO$_2$ layers with the $90^\circ$ angles of the bonds, see Ref.~\onlinecite{khaliullin2008origin}.}
Such processes are allowed in the layers, because the angles of the Co-O-Co bonds are about $90^\circ$.
Moreover, these processes are plausible, because the Hund coupling between the elections within the $d$-shells substantially reduces the energy $\approx \Delta-2\,J_\text{H}\approx 300~\text{meV}$\cite{Khaliullin2005,khaliullin2008origin} for the corresponding transitions of the Co$^{3+}$ ions across the crystal gap.
\footnote{For the nearest $d$-sites in the CoO$_2$ slabs with the $90^\circ$-bonds Co-O-Co, the magnitude $t'$ of the parameters of the electron hopping between the $t_{2g}$ and $e_g$ states is close to the value of the hopping integral $t$ between the $t_{2g}$ states of the neighbors: $t'\gtrsim t\approx 0.2~\text{eV}$, see Ref.~\onlinecite{Chaloupka2008}.}
Then the holes become ``dressed'' by clouds of the HS-spin excitations to the $e_g$-orbital.
When such polarons propagate throughout the $d$-lattice they mix the HS and LS states of Co$^{3+}$ ions.\cite{khaliullin2008origin}
However, the polaron scenario contradicts the picture of the disproportionation, where the significant part $\approx25\%$ of the  Co1$^{3+}$ sites are not affected by the hole hopping, remaining in the non-magnetic state according to the NMR experiments.
Also this approach does not take into account the effect of ordering of the sodium ions in the cobaltates.

Thus, the electron disproportionation with the peculiar KSL-structure was not obtained by the previous theories, and the physics of the cobaltate metallic state remained controversial.
Following the LS scheme, Koshibae and Maekawa put forward a model, where the charge disproportionation in the layer resulted from the orbital states of $t_{2g}$-holes, having assumed that a splitting of their $t_{2g}$ triplets by the crystal distortions is negligible.\cite{MaekawaKoshibae2003}
In this theory, an energy dispersion of the non-interacting holes was found to possess four degenerated branches each corresponding to one of the four separate kagom\'e sublattices.
Each KSL is formed by its own sequence of the hole hoppings between the respective orbitals of the $t_{2g}$ triplets.
It was suggested in Ref.~\onlinecite{MaekawaKoshibae2003} that the kagom\'e state of the holes is hidden in the LS excitations of the layers in the vicinity of the Fermi level, and the commensurate order of the Na$^+$ ions could stabilize it.
In subsequent calculations it was established, that when the coupling with such order is taken into account, the LS states with hole disproportionation indeed appear, possessing charge/spin patterns of various types (the zig-zag, striped, honeycomb, and hexagonal) which depend on the presumed Na arrangement.\cite{Lee2004,Li2005,Indergand2005,Lee2005,Lee2006,lysogorskiy2012ab}
However the ordering of Na$^+$ ions drastically changes the electronic low-energy states,\cite{Lee2004} and the KSL structure expected for the phase $x=2/3$ was not obtained in the LS-states of the lattice even with the realistic sodium structure.\cite{lysogorskiy2012ab}

The aim of the present study is to investigate conditions for the origin of a metallic state with the KSL-structure of the $d$-electron charge and spin disproportionation in cobaltates, employing the \textit{ab-initio} calculations for the system with ordered Na ions.
Out approach to the problem takes into account the interaction between the itinerant holes and the \emph{HS excitations} of the lattice.
In Sec.~\ref{sec:Experimental} we study the magnetic susceptibility of the high-quality single-crystal  Na$_{2/3}$CoO$_2$ compound to clarify a possibility of existence of the HS-states.
Then in Sec.~\ref{sec:abinitio}, employing the GGA+U approximation, we calculate the collective electronic states and magnetic/charge structure of the cobalt layers for the $x=2/3$ crystal phase with the realistic ordering of the sodium ions.
We analyze the non-uniform LS and HS states of the Co lattice in a dependence on the parameter $U-J$ of the electronic correlations within the $d$-shells and compare the energy of the states.
In Sec.~\ref{sec:discussion} we discuss the electron organization of the obtained metallic states with the charge/magnetic disproportionation.

\section{Experimental}
\label{sec:Experimental}

\begin{figure}[tbp]
\includegraphics[width=1.0\linewidth]{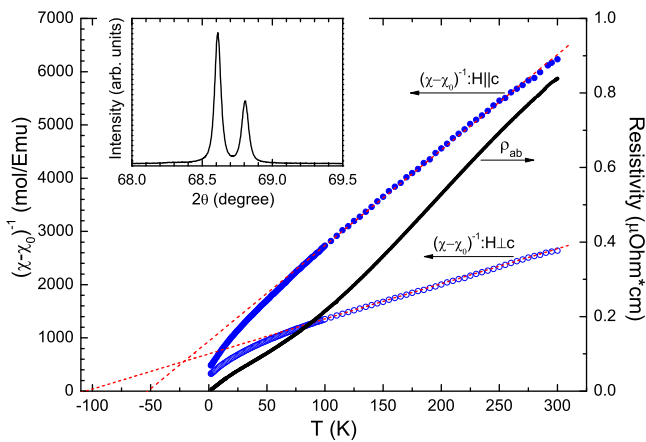}
\caption{(Color online) Temperature dependencies of the inverse spin susceptibility $(\chi-\chi_{0})^{-1}$ of the Na$_{2/3}$CoO$_{2}$ single crystal in a $1$\,T applied magnetic field and the in-plane resistivity $\rho_{ab}$ of the same crystal. The inset shows the part of the x-ray spectra with the ($008$) Bragg peaks, which permits phase identification and ensures the absence of phase mixing (the double peak structure corresponds to two Bragg peaks associated with the Cu K$\protect\alpha _{1}$ and K$\protect\alpha _{2}$ radiation lines).}
\label{FigNa067Exp}
\end{figure}

A series of sodium cobaltates Na$_{x}$CoO$_{2}$ with $x\approx 0.8$ were grown by the floating zone technique.\cite{mukhamedshin2014complex} Using an electrochemical Na de-intercalation method, we reduced the sodium content in the as-grown crystals down to that of the pure phase with $x=2/3$. In the inset of Fig.~\ref{FigNa067Exp} the part of the x-ray spectra with the ($008$) Bragg peak of obtained single crystal is shown. The position of this peak corresponds to $x=2/3$ \cite{AlloulEPL2008,mukhamedshin2015order} and unambiguously proves phase purity of our samples.

The temperature dependence of magnetic susceptibility $\chi=M/H$ is measured with Quantum Design MPMS-XL in the temperature range $1.8$-$300$\,K with magnetic fields of $1$\,T applied parallel and perpendicular to the $c$ axis, respectively. In the high temperature region $100$-$300$\,K experimental curves were fitted by the function
\begin{equation}
\chi = \chi_{0}+\chi_{s}(T) = \chi_{0}+\frac{C}{T-\theta}.
\label{eq:Curie-Weiss}
\end{equation}%
In Eq.~(\ref{eq:Curie-Weiss}) the first term $\chi_{0}$ contains temperature independent contributions to the susceptibility: the diamagnetic component, Pauli, and orbital paramagnetism. The second term is the temperature dependent contribution to the spin susceptibility $\chi_{s}$ which follows the Curie-Weiss law with Curie constant $C$ and Curie temperature $\theta$. In the main panel of Fig.~\ref{FigNa067Exp} the temperature dependencies of the inverse spin susceptibility $(\chi-\chi_{0})^{-1}$ of the Na$_{2/3}$CoO$_{2}$ single crystal are presented for the two orientation of the applied field $H\parallel c$ and $H\perp c$. As one can see, the spin susceptibility of the Na$_{2/3}$CoO$_{2}$ single crystal is very anisotropic and are characterized by different values of $C$ and $\theta$ in the two directions. The origin of such anisotropy is unclear.

For a paramagnet with effective magnetic $d$-moments $p_\textrm{eff}$ the Curie constant $C=\frac{nN_{A}p_\textrm{eff}^2}{3k_{B}}$ where $n$ is the fraction of cobalt ions having a magnetic moment. The $^{59}$Co NMR experiments have clearly proved $25\%$ of cobalt sites in the Na$_{2/3}$CoO$_{2}$ phase being very close to the non-magnetic Co$^{3+}$ state, and only $75\%$ of the cobalt ions (Co2 sites) have magnetic moments.\cite{Platova2009} Therefore the effective moment per Co2 site is equal to $1.27(1)$~$\mu_{B}$ [$C=0.152(3)$~emu K/mol] and $0.77(1)$~$\mu_{B}$ [$C=0.056(2)$~emu K/mol] in the case of $H\perp c$ and $H\parallel c$, respectively. However a relevance of such analysis, which is valid for \textit{localized moments}, to the susceptibility of sodium cobaltates, which are the systems with itinerant magnetism, is doubtful. Nevertheless it indicates  that the magnetic moments of cobalt ions in the sodium cobaltates could be larger than that of an isolated spin $\frac{1}{2}$. 

Also in the main panel of Fig.~\ref{FigNa067Exp} the temperature dependencies of in-plane resistivity $\rho_{ab}$ of the same Na$_{2/3}$CoO$_{2}$ phase single-crystal is shown. It was measured with PPMS system (Quantum Design) and AC transport measurement option in a zero magnetic field and at frequency $103$\,Hz. The Van der Pauw measurements method was used and contacts with low resistivity were made by attaching gold wires with silver epoxy to the corners of single crystal with $2\times2\times0.11$\,mm size. As one can see in Fig.~\ref{FigNa067Exp} the resistivity of Na$_{2/3}$CoO$_{2}$ phase has a ``metallic'' dependence on $T$, being almost linear within the range $100$-$300$~K. It is similar to the data reported in Ref.~\onlinecite{Foo2004} for the $x=0.71$ sample.

\section{\emph{Ab initio} study}
\label{sec:abinitio}
\subsection{Method}

In our study we calculate the electronic states of the sodium cobaltate from the first-principles, using the spin-polarized density functional theory~\cite{kohn1965self} (DFT) with the generalized gradient approximation (GGA).
The GGA employs the functional form of the Perdew-Burke-Ernzerhof exchange correlation revised for solids (PBEsol).~\cite{PBEsol}
The Coulomb interaction between the electrons and ionic cores is taken into account using the projector augmented-wave method (PAW),~\cite{BlochlPAW} which was implemented in the Vienna Ab Initio Simulation Package~\cite{kresse1996efficient} (VASP 5.2.12) (a part of the MedeA software package~\footnote{MedeA version 2.16. MedeA is a registered trademark of Materials Design, Inc.,Angel Fire, New Mexico, USA.}).
The plane-wave cutoff energy is $500~\text{eV}$.
The Brillouin zone is sampled by the $5 \times 5 \times 5$ mesh, containing $125$ points including the $\Gamma$ point  $(0,0,0)$ in the reciprocal space.
Correspondingly the  linear density of the mesh is one point per $\approx 0.150$ \AA$^{-1}$.
Again, the convergence criteria $10^{-6}$ eV is chosen for the electronic energy.

We treat the electronic correlations in the $3d$-shells of cobalt within the rotationally invariant GGA+U approach, introduced by Dudarev \textit{et al.},~\cite{Dudarev} where the effective Hubbard parameter $U_\textrm{eff}=U-J$ is meaningful. The subscript ``eff'' is hereafter skipped for a briefness.
The electronic states were calculated in a dependence on $U$ in the range from $0$ to $6$\,eV.
For the cobaltates the relevant value of $U$ is about $5$\,eV, see Ref.~\onlinecite{Hasan2004a}.

In our calculations we employ the lattice with complete crystallographic cell of the  Na$_{2/3}$CoO$_2$ compound, containing $264$ atoms, which was previously deduced from the  NMR/NQR experiments and verified by XRD Rietveld measurements in Ref.~\onlinecite{AlloulEPL2008}.
To optimize the total energy of the model, we ``relax'' the lattice using conjugated gradient method:
sites are slightly shifted towards their equilibrium positions in such a way that a maximal force affecting the sites becomes less than  $0.01~\text{eV}$\AA$^{-1}$.

To obtain the different charge/magnetic collective states of the given model, in the calculations we examined the respective trial (initial) magnetic moments of the eight non-equivalent cobalt positions in the crystal.
Both the LS and HS configurations of the Co ions were taken into account in our study.
In particular, to get the hexagonal LS state of the lattice [Fig.~\ref{fig:Kagome}(c)] the useful initial assumptions were found to be $0$\,\muB for the Co1 sites and $0.5$\,\muB for the Co2 positions,
whereas for the HS-KSL state [Fig.~\ref{fig:Kagome}(d)] those were $0$\,\muB and $1.5$\,\muB respectively.

\subsection{Electronic states}
\label{sec:ElectronicStates}

\begin{figure}[t!]
	\centering
	\includegraphics[width=0.95\linewidth]{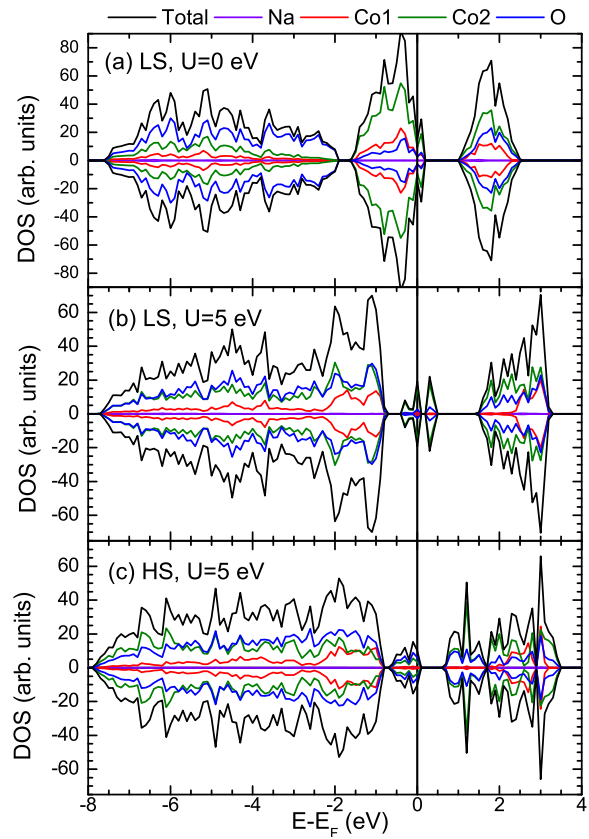}\\
	\caption{(Color online) \label{fig:DOS} Spin resolved projected density of electronic states of Na$_{2/3}$CoO$_2$ crystal, plotted versus the energy: (a) LS state at $U=0$\,eV, (b) LS state at $U=5$\,eV, and (c) HS-KSL at $U=5$\,eV. DOS per atoms for the spin-up (spin-down) states are given above (below) the horizontal reference lines. The solid vertical line indicates the Fermi level.}
\end{figure}

\begin{figure}[tbp]
\includegraphics[width=1.0\linewidth]{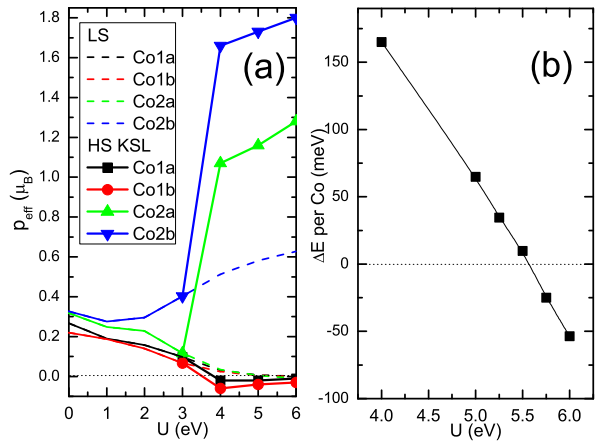}
\caption{\label{fig:dEISLS} (Color online) (Color online) Magnetic moments and energy of the system, calculated in the dependence on the parameter of the electronic interaction within the $d$(Co)-shells. (a) Moments $\mu$ of the Co sites. (b) Difference between the energy of HS-KSL and LS states estimated per Co ion.}
\end{figure}

\begin{table}[t!]
	\caption{\label{tab:tab1}
Calculated magnetic moments of the nonequivalent Co sites (in units of $\mu_\mathrm{B}$).\footnote{The moments were estimated by integrating the electronic spin density over the sphere with radius $1.16$~\AA~centered at the respective Co sites.}
Average values are given in the last column.}
	\def\arraystretch{1.5}
	\begin{adjustbox}{center}
		\begin{ruledtabular}
			\begin{tabular*}{\columnwidth}{cccccc}
				Model state  & {Co1a}& {Co1b}& {Co2a}& {Co2b} & {$\langle\mu\rangle$ per Co}\\
				\hline
				{LS, U = 0\,eV} & {0.174} & {0.254} & {0.299} & {0.344} & {0.297}  \\
				\hline
				{LS, U = 5 eV} & {0.016} & {0.018} & {0.020} & {0.621} & {0.320}  \\
				\hline
				{HS-KSL, U=5 eV} & {0.015} & {0.046} & {1.233} & {1.879} & {1.254} \\
			\end{tabular*}
		\end{ruledtabular}
	\end{adjustbox}
\end{table}

Now we consider the electronic states, calculated at $T=0\,\text{K}$ within the GGA+U approximation for the system, which possesses the realistic crystal structure of the Na$_{2/3}$CoO$_2$ compound with the ordered sodium ions.
In such approach the non-uniform charge/magnetic states of the electrons appear when their correlations are enhanced within the $d$-shells.
When the correlations are weak, the interaction of the charged $t_{2g}$-holes with the non-uniform crystal potential introduced by the sodium order is not efficient: at $U=0$ the model has metallic state, being almost \emph{uniform} in the cobalt planes, which has a slight FM instability of the itinerant character.\cite{moriya2012spin}
The hole conduction $t_{2g}$-band has a large width $W \approx 1.5~\text{eV}$ and is separated from the upper $e_g$-band by the gap $\approx 1~\text{eV}$.
The energy dependence of the corresponding electronic density of states (DOS) is presented in Fig.~\ref{fig:DOS}(a).
The calculated state resembles that which was obtained in the cobaltates with the uniform sodium ion distribution within the LSDA approximation.~\cite{Singh2000}

When the correlations in the $d$-shells become stronger, qualitatively new metallic states, being heterogeneous in the Co sublattice, emerge in our model.
The magnetic moments of the $d$-shells serve as indicators of the redistribution of the $d$-electron spin and charge density between the Co ions in their triangular lattice.~\cite{AlloulEPL2009,Lee2004,Li2005,Lee2005,lysogorskiy2012ab}

In the previous study we have reported about the LS ``phase'' of the electrons,~\cite{lysogorskiy2012ab} see Fig.~\ref{fig:Kagome}(c). This non-uniform LS state emerges upon increasing $U$:
when $U\gtrsim 2~\text{eV}$ the moments of Co2b sites  increase progressively together with a marked decreasing of the Co1 and Co2a site moments, see Fig.~\ref{fig:dEISLS}(a).
When $U$ exceeds 4\,eV the LS-state becomes strongly heterogeneous.

In the present investigation we have found another state, called HS-KSL phase: the moments of the $d$-sites are enhanced [see Fig.~\ref{fig:dEISLS}(a) and Table~\ref{tab:tab1}] and the KSL in triangular cobalt lattice is formed, see Fig.~\ref{fig:Kagome}(d).
Initially, the energy of the HS-KSL state is larger by $\approx150$\,meV than the energy of the LS-phase, see Fig.~\ref{fig:dEISLS}(b).
Then, the difference between the energies of these states decreases upon increasing $U$, and the HS-KSL phase becomes the ground state of the system when $U\gtrsim5.4~\text{eV}$.

Energies of these two collective states are close, and they compete with each other in the system with the realistic $U\approx5~\text{eV}$.

In both phases the Co1 sites become almost nonmagnetic at realistic $U$, see Figs.~\ref{fig:Kagome}(c),~\ref{fig:Kagome}(d) and~\ref{fig:dEISLS}(a).
This implies the states of these ions to be close to the spinless state $t_{2g}^6$ (Co$^{3+}$):
magnetic holes Co$^{4+}$ avoid the Co1 positions due to sodium ions being close to these cobalt sites,~\cite{AlloulEPL2009} see Fig.~\ref{fig:Kagome}(a).
However in the LS-phase holes also avoid the Co2a sites, remaining nonmagnetic as well, see  Fig.~\ref{fig:dEISLS}(a).
Instead, the holes tend to the Co2b sites, forming the magnetic hexagons in the triangular lattice [Fig.~\ref{fig:Kagome}(c)].
The magnetic moments of these sites are small, so that their spins $S < 1/2$, see Table~\ref{tab:tab1}.
It means that Co2b ions are mostly in their LS states. 
The organization of the collective LS-state resembles the droplet electronic phase of the multi-orbital Hubbard model on the triangular lattice.~\cite{Fevrier2015}
In our approach the hexagonal hole ``droplets'' are likely ``trapped'' by the sodium potential, arranging them in the form of triangular super-lattice with doubled period.

On the contrary, in the HS-KSL phase Co2a sites are also magnetic along with Co2b sites, thereby the hexagons become connected resulting in the kagom\'e pattern, see Fig.~\ref{fig:Kagome}(d).
Such state of the model is consistent with the magnetic and charge structure, established in the cobalt sublattice of the Na$_{2/3}$CoO$_2$ compound using the NMR/NQR technique,\cite{AlloulEPL2009,Platova2009} cf. Figs.~\ref{fig:Kagome}(b) with \ref{fig:Kagome}(d).
In our calculations the average magnetic moment on the Co2 sublattice $p_\textrm{eff}=1.66$\,$\mu_\textrm{B}$ corresponds to the enlarged average spin $S=0.83$, indicating an existence of the HS states in the KSL.

Both the LS and HS-KSL phase has the A-AFM structure with inplane/interplane FM/AFM ordering of the $d$-moments.
The GGA+U approximation is known to overestimate the FM interaction within the CoO$_2$ layers.
For instance, it results in the band ferromagnetism for the cobaltates with $x<0.5$, which experimentally has the in-plane AFM correlations.~\cite{zhang2004doping}
Nevertheless, both calculated states correctly represent the character of the correlations between the $d$-moments in the disproportionation domain $0.65 \lesssim x \lesssim 0.8$ of the phase diagram of Na$_x$CoO$_2$.
Again, both the LS and HS-KSL states are found to be metallic.
In particular, the narrow band with width $\approx0.5$\,eV is formed at the Fermi level instead of the initial wide conduction band, cf. Figs.~\ref{fig:DOS}(b) and (c) with (a).
This implies that heterogeneity of the many-particle states and enhanced electronic correlations appear concurrently.

The decrease of the energy gap between the HS-KSL and hexagonal phase and the formation of HS states of the sites of KSL go hand-in-hand upon increasing $U$, cf. Fig.~\ref{fig:dEISLS}(a) and (b).
This suggests that the energy of the HS-KSL phase reduces towards that of the LS phase due to a gain in the energy of the electron Hund interaction within the $d$(Co2)-shells.
A likely origin of such dependence on $U$ could be the following:
the intersite hopping of electrons with arbitrary spin states results in local magnetic fluctuations in the metal.\cite{pines1990theory}
Then, within the $d$-shells these spin fluctuations oppose the spin arrangement due to the FM Hund coupling between the electrons, which tends to maximize the on-site total spins.
However, upon increasing $U$ both the on-site and inter-site correlations between the itinerant particles become enhanced.\cite{gebhard2003mott}
The correlations thereby help the Hund mechanism to overcome and increase the $d$-moments in the KSL by attenuating the arbitrary spin fluctuations in the metal to some extent.

We have not found a ``LS kagom\'e'' phase, possessing both the small spins $S<1/2$ of the Co2 ions and the KSL disproportionation of lattice.
This implies that the LS $t_{2g}$-states of the Co2a sites that play a role of bridges for the hole hopping between the hexagons to form the KSL-state, are not efficient by itself.
However, the obtained HS-KSL state provides the possibility for the holes to bridge the hexagons and to form the KSL, involving the HS $e_g$-states of the sites.

\section{Discussion}
\label{sec:discussion}

The NMR measurements detected the marked violation of the axial symmetry of the $d$-shells of the magnetic  Co2 sites, which appears in the cobalt planes simultaneously with the electron disproportionation.\cite{Platova2009}
Within the LS-approach, the  Co$^{4+}$ ($t_{2g}^5$) holes hop between the $t_{2g}$ on-site states in the lattice of the inert  Co$^{3+}$ ($t_{2g}^6$) ions.
The chemical compression of the CoO$_2$ layers along the $c$-axis direction yields the trigonal deformation of the  CoO$_6$ octahedra, and the on-site $t_{2g}$-triplet states become splitted into the ground $e'_g$ and excited $a_{1g}$  orbital states with the axially symmetric $a_{1g}$ wave functions extended along the $c$-direction.\cite{Landron2006} Then, to obtain the asymmetry of the local $d$-shells, one has to take into account transitions of the \textit{holes} into their anisotropic $e'_g$ orbital states along with the $a_{1g}$ orbitals.
However, the \emph{ab-initio} calculations for the cobaltates indicate that the energies $\approx300~\text{meV}$ of these LS $e'_g$-excitations of the holes\cite{Landron2006} match the typical energy $\approx250-300~\text{meV}$ of excitations of the Co$^{3+}$ ions from their spinless to HS $e_g$-states in the transition metal oxides.\cite{Nekrasov2003,Chaloupka2008}
This suggests, that the theory of charge disproportionation in the cobaltates should take into account both the LS-states of the holes and HS-states of their neighboring $d$-ions.
Thus, the anisotropy of the $d$-shells, accompanying the appearance of the kagom\'e state in the experiments, just implies that the coupling between the holes and the HS-states of the $d$-lattice are relevant for this phenomena.
We shall now discuss the formation of the KSL disproportionation within our approach.

As shown in Sec.~\ref{sec:ElectronicStates}, when the LS states become heterogeneous the itinerant holes are almost bound within the Co2b hexagons, see Fig.~\ref{fig:Kagome}(c).
The holes localized in these hexagons reduce the energy of mutual on-site Coulomb repulsion between them, correlating their spins ferromagnetically.
This manifests itself as a small increase of the magnetic moments of Co2b sites upon increasing $U$, see Fig.~\ref{fig:dEISLS}(a).
However, the localization of itinerant holes increases a kinetic energy of the system.
\footnote{At present, a mechanism of the formation of the hexagonal LS state of $t_{2g}$-holes, obtained in our \emph{ab-initio} study of the cobaltates, is not evident and could be investigated in a further research within a simplified model. The hexagonal arrangement of the holes is likely related with their ``kinematic interaction'', predicted in the Ref.~\onlinecite{MaekawaKoshibae2003}. Then, such organization of the hole states reflects a peculiarity of the $t_{2g}$-orbital space of the compressed layers CoO$_2$, where the trigonal deformations of the octahedra splits the on-site $t_{2g}$-triplets of the Co planes. Again, these states appear in the non-uniform crystal field introduced by the sodium ion order. In particular, the electric field of Na1$^+$ cations pushes the holes out of the Co1b sites, the centers of the hexagons.}

The hybridization between the states of the holes in $t_{2g}$ orbitals and HS $e_g$-states of the lattice~\cite{Khaliullin2005} promotes the tunneling of the holes from the Co2b hexagons to the $d$-shells of their nearest neighbors: this decreases the kinetic energy.
To this end, among the nearest  Co$^{3+}$ ions the holes prefer to involve the $e_g$-states of  Co2a sites rather than those of Co1 sites, see Fig.~\ref{fig:Kagome}(d). For both the Co2a and Co2b ions, the energy of their on-site excitations from the lower $t_{2g}$-states to the upper HS $e_g$-states is reduced due to the strong violation of the trigonal symmetry $C_{3v}$ of the crystal environment in these atomic positions. This symmetry breaking results in the splitting of the $e_g$ doublets of the Co2 sites.

On the contrary, for the Co1 ions the symmetry $C_{3v}$ is not violated, and do not affect the energy gap for the on-site $e_g$-excitations.
Again, the Coulomb repulsion from the  Na$^+$ ions, being particularly close to the Co1 positions, suppress the hopping  of the holes, occupying the  Co2 sites, to the Co1 sites, cf. Fig.~\ref{fig:Kagome}(a) with \ref{fig:Kagome}(d).
As the result, the hybridization between the states of the holes and Co1 sites turns out to be small.
This explains the nonmagnetic states of the Co1 sites,  resembling the LS spinless state $t_{2g}^6$ of the Co$^{3+}$ ions, in the $d$-lattice.

Therefore the itinerant holes, interacting with the HS $e_g$-states of the cobalt ions, form the metallic HS-KSL state extended over the triangular lattice, which \emph{competes} with the almost localized hexagonal LS state.

Now we consider the plausible mechanism, which results in the considerable splitting of the $e_g$(Co2)-doublets and weakly affects the doublets of the Co1 sites.
The interaction, relevant for such effect, could be the coupling between the $e_g$-electrons of Co sites and the deformation of the octahedra of the oxygen anions, ligands of the corresponding $d$-ions

In particular, in our model the optimization of atom positions reveals a peculiar distortion of the CoO$_2$ layers, which is specific for the Co1 and Co2 sites.
Namely, we have found that the Co2-O$_6$ octahedra could possess substantial deformation modes with $E_g$-symmetry, $Q_\epsilon$ ($x^2-y^2$) and $Q_\theta$ ($2z^2-x^2-y^2$),~\cite{bersuker2006jahn} whereas these deformations are not developed in the case of  Co1-O$_6$ octahedra.
Broadly speaking, such deformations exist in the every electronic states, calculated as with $U=0$ and so with its larger values.
However in the HS-KSL phase the $E_g$-modes are by order of magnitude stronger than those evaluated in the other states of system.
To be specific, the magnitude reaches the substantial value $\approx 0.2~\text{\AA}$ at $U=5~\text{eV}$, i.e., about $10\%$ of the Co-O bond length $\approx 1.9$~\AA.
Such strong $Q_{\epsilon,\theta}$ deformations of the Jahn-Teller (JT) form are typical for the systems with active $e_g$-electrons, e.g., for the manganites.~\cite{KilianKhaliullin1998}

Cooperative JT distortions polarize the $e_g$-wave functions $x^2-y^2$ and $2z^2-x^2-y^2$, extended along the  Co-O bonds, thereby splitting these doublets.
In turn the $t_{2g}$ states, $xy$, $yz$, and $zx$, weakly interact with the JT lattice modes, because lobes of these orbitals are directed aside from the anionic sites.\cite{Khaliullin2005}
As a result, the crystal gap between the $e_g$ and $t_{2g}$ states decreases together with the energy of the HS states of  Co2 ions containing the $e_g$ electrons.
The interaction between the $e_g$-electrons and the crystal deformation thereby promote the reducing of energy of the HS-KSL state with respect to the LS state.
In the collective LS states this interaction is relatively weak, because they mostly have the $t_{2g}$-orbital type.

Surprisingly, the Co2 positions, forming the kagom\'e lattice, are disposed along sides of the triangles of Na2 atoms, cf. Fig.~\ref{fig:Kagome}(d) with~\ref{fig:Kagome}(a).
Thus, the deformation modes $Q_{\epsilon,\theta}$, promoted by the interaction with the HS $e_g$-states distributed over the KSL by the hole motion, could be triggered by the chemical tension, accumulated in the  Co2-O$_6$ octahedra at the border of the sodium clusters.
This suggests the explanation of the pronounced electron-lattice effect observed in the HS-KSL state.
Compressing the CoO$_2$ layers from the opposite sides, the triangular Na-blocks tend to skew the octahedra in-questions situated between their edges.
On the contrary, the  Co1-O$_6$ octahedra are mainly squeezed along the direction of the $c$-axis, because the Co1 sites are placed about the centers of the Na2-triangles and about the lone ions Na1, see Fig.~\ref{fig:Kagome}(a).
Then, the related octahedra distortion does not substantially violate the trigonal symmetry of the crystal environment at the Co1 positions, and, thus, does not considerably affect the $e_g$-states of these sites.

When the sodium ions are distributed randomly above and below the cobaltate layers, the GGA+U approximation results in uniform HS states of the  CoO$_2$ layers, which are separated from the corresponding uniform LS ground-state by the large energy gap.
This implies that the uniform HS state could unlikely appear in the metallic cobaltates.\cite{shorikov2011role}

In our approach, the interaction with the ordered sodium layers ``reconstructs'' the electron wave functions and the energy spectrum of the system around the Fermi level at realistic $U$.
As a result, \emph{both} the LS and HS states become heterogeneous and competing states, being almost degenerate in their energy.
This is evident in Fig.~\ref{fig:dEISLS}(b): when $U\approx 5~\text{eV}$ the energy of the HS-KSL state is larger by only $60~\text{meV}$ than the energy of the LS state, and already at the same values $U\gtrsim 5.5~\text{eV}$ the HS state has the less energy as compared with the LS one.
The competing LS and HS states could be intermixed by collective fluctuations, which should become intensive in the critical region.~\cite{sachdev2007quantum}

One of the mechanisms, introducing such low-frequency fluctuations, could be the electron-phonon coupling, which provides a creation and annihilation of virtual phonons in the lattice.~\cite{migdal1958interaction}
In the cobaltates a typical frequency of the phonons with $E_{1g}$-symmetry, interacting with the $e_g$-orbitals, is just $\approx60~\text{meV}$.~\cite{Donkov2008}
Another mechanism mixing the LS and HS states could be the inter-site electron hopping within the $90^\circ$ Co-O-Co bonds.~\cite{Khaliullin2005}
This results in the spin/orbital correlations between the $d$-sites,~\cite{daghofer2006magnetic, chaloupka2007spin, khaliullin2008origin} which are underestimated in the GGA+U approximation.

Again, the HS-states have a larger entropy than the LS-states, and when the temperature increases, these states could progressively emerge in the state of system due to thermal excitations.\cite{Yamaguchi1996}
Another origin of the appearance of HS-states could be the thermal expansion of lattice upon increasing $T$, which reduces the energy of the $e_g$-excitations.
In particular, the expansion enlarges the distance between the $d$-ions and ligands, thereby reducing the hybridization between their wave functions. Consequently, the splitting of the $d$-levels of Co sites decreases. For instance, it was established that such mechanism is indeed able to stabilize the uniform HS-state,\cite{Nekrasov2003} which was experimentally observed in the perovskite LaCoO$_3$ at $T<130~\text{K}$.\cite{Yamaguchi1996}

To evaluate the influence of thermal expansion on the energy difference between the HS-KSL and LS phase, we have correspondingly enlarged their crystal cells, calculated at $T=0~\text{K}$ in our model.
Namely, the initial cells with the respective volumes $V_{0,\text{LS}}=2618\,\text{\AA}^3$ and $V_{0,\text{HS-KSL}}=2717\,\text{\AA}^3$
were evenly expanded to reach the cell volume $V_T=2730\,\text{\AA}^3$, established experimentally in Na$_{2/3}$CoO$_2$ at $T=300~\text{K}$.
After this, the energies of HS-KSL and LS electronic phases were calculated for expanded lattices at $U=5~\text{eV}$.
We have thereby obtained that such expansion indeed markedly reduces the initial energy gap to the value $37~\text{meV}$ per Co site at room temperature.
A comprehensive analysis of such temperature effect demands experimental data on the crystal parameters in a dependence on $T$.

The above mechanisms qualitatively explain the progressive increase of the magnetic moments of Co ions upon increasing $T$, which was deduced from the measurements of magnetic susceptibility of the system, see Sec.~\ref{sec:Experimental}.
A careful description of the temperature dependence of the metallic state of the cobaltates demands an explicit account of the effects of quantum and thermal fluctuations.
Such research is beyond the scope of the present GGA+U approach and is remained for future studies.

In the picture of the competing hexagonal LS and HS-KSL states the Co1 sites remain inert in the triangular $d$-lattice, cf. Fig.~\ref{fig:Kagome}(c) with (d).
The holes are related with the Co2 KSL, where the more and less magnetic  Co2b and Co2a sites are formed, respectively.
Such structure conforms to the pattern of disproportionation of the $d$-ion states deduced from the NMR/NQR measurements in Na$_{2/3}$CoO$_2$,\cite{AlloulEPL2009} see Fig.~\ref{fig:Kagome}(b).
The suggested heterogeneous HS state substantially differs from the polaronic HS state proposed in Ref.~\onlinecite{khaliullin2008origin}, which runs over the whole sites of the triangular lattice doped by the itinerant holes.

\section{Conclusion}

In the present work we have investigated an origin of the disproportionation state of the metallic lamellar cobaltate Na$_{x}$CoO$_2$, i.e., substantially  heterogeneous distribution of the charge and magnetic density of the electrons among the d-ions in the triangular cobalt lattice.
To this end, we have calculated the electronic states and their energy spectrum from first principles for the system possessing the realistic crystal structure.
The states have been analyzed in a dependence on the strength of the electron correlations between the electrons at the $d$-sites.
The model takes into account the correlations, caused by the electron Coulomb repulsion and their Hund spin coupling within the $d$-shells, treating this on-site interaction with the GGA+U approximation.
To be specific, we have calculated the magnetic and charge structure of the Co lattice of the representative crystal phase with $x=2/3$ of the cobaltate.
In our approach, the heterogeneity of the $d$-states is triggered by the clustering of the Na$^+$ cations at the Na1/Na2 position of the sodium layers.

The heterogeneity was obtained to increase upon increasing the parameter of the on-site correlations.
First, the electronic LS state with the small $d$-moments develops, and then the HS-KSL state with the increased moments emerges.
When the heterogeneity becomes strong these states remain metallic.

We have established, that the HS-KSL state of the $d$-lattice combines the inherent features of the experimental state of the cobaltate.\cite{AlloulEPL2009,Platova2009}
It is the disproportionation state, which contains the non-magnetic and magnetic $d$-sites with the ratio $1:3$ between their portions.
In this state the kagom\'e structure appears, which contains the itinerant holes.
The KSL is formed by the more magnetic Co2b sites assembled in the hexagons, connected by the less magnetic  Co2a sites, see Fig.~\ref{fig:Kagome}(d).
On the contrary, in the LS state the kagom\'e structure splits into the separate hexagons, see Fig.~\ref{fig:Kagome}(c).
We have found that the energies of the HS-KSL and LS states become close to each other at the realistic values of the correlation parameter. Their competition could therefore provide the KSL-structure of the disproportionation, revealed in the compound Na$_{2/3}$CoO$_2$ in Refs.~\onlinecite{AlloulEPL2009,Platova2009} with the NMR/NQR technique.

We have pointed out, that the enhanced moments obtained in the HS-KSL state imply the participation of the HS-states of Co $d$-orbitals in the formation of the KSL structure.
Imprints of the enhanced $d$-moments of the Co2 ions have indeed been found in our data on the magnetic susceptibility measurements of the single-crystal sample  Na$_{2/3}$CoO$_2$.

The present study thereby suggests the key to understand the physics of the intricate metallic states of the cobaltate in the range $0.65<x<0.8$, where the enhanced many-particle correlations and the disproportionation of the $d$-electrons emerge concurrently.
In the given approach the relevant correlations are related both with the on-site Coulomb repulsion and Hund coupling between the $d$-electrons.

\section{Acknowledgements}

This work was funded by the subsidy allocated to Kazan Federal University for the state assignment in the sphere of scientific activities and also by the RFBR under Project 14-02-01213a. The work was supported by Russian Government Program of Competitive Growth of Kazan Federal University. Authors are grateful to A.\,V.~Dooglav for his help with the manuscript preparation.

%

\end{document}